# Phase space topology of four-wave mixing reconstructed by a neural network

ANASTASIIA SHEVELEVA,[1] PIERRE COLMAN,[1] JOHN. M. DUDLEY,[2] CHRISTOPHE FINOT[1,*]

[1] *Laboratoire Interdisciplinaire Carnot de Bourgogne, UMR 6303 CNRS-Université de Bourgogne-Franche-Comté, 9 avenue Alain Savary, BP 47870, 21078 Dijon Cedex, France*
[2] *Institut FEMTO-ST, Université Bourgogne Franche-Comté CNRS UMR 6174, Besançon, 25000, France.*
\* *Corresponding author: christophe.finot@u-bourgogne.fr*

**The dynamics of ideal four-wave mixing in optical fiber is reconstructed by taking advantage of the combination of experimental measurements with supervised machine learning strategies. The training data consist of power-dependent spectral phase and amplitude recorded at the output of a short segment of fiber. The neural network is able to accurately predict the nonlinear dynamics over tens of kilometers, and to retrieve the main features of the phase space topology including multiple Fermi-Pasta-Ulam recurrence cycles and the system separatrix boundary.**

The nonlinear Schrödinger equation (NLSE) is one of the seminal equations of science, describing wave evolution in a dispersive medium subject to an intensity-dependent nonlinear response. Because of their low attenuation, optical single mode fibers are an ideal testbed to investigate the complex dynamics of the NLSE-ruled systems. Amongst the key physical process involved in the NLSE is nonlinear four-wave mixing (FWM) associated with dynamical energy exchange between discrete evolving frequency components [1]. An NLSE configuration of particular interest is the degenerate case when a single frequency pump generates only two sidebands of upshifted and downshifted frequency so that the nonlinear dynamics can be reduced to a system of three coupled differential equations [2]. This is the simplest configuration with which to observe FWM, and its understanding is the key to interpret more complex phenomena in nonlinear optics.

Although this canonical FWM system has been the subject of a number of previous theoretical and numerical studies, it is notoriously difficult to implement in practice due to the emergence of additional frequency components, and because of the impact of residual loss [3]. We have recently demonstrated an experimental solution to this problem which limits the generation of higher-order sidebands so that the dynamics can be accurately described by a set of three coupled equations. In particular, we have developed a system where iterated (programmed) initial conditions are sequentially reinjected into an optical fiber [4], allowing experimental observation of the full dynamical phase-space topology, including multiple Fermi-Pasta-Ulam recurrence cycles, stationary wave existence, and the system separatrix boundary that distinguishes qualitatively different regimes of evolution.

A particular advantage of these experiments is the ability to conveniently explore the full dynamical phase space of pump and sideband evolution through the use of programmable injected amplitude and phase. In this Letter we show how this particular feature of the system can be combined with machine learning techniques [5-7] to train a neural network to accurately model and predict the underlying FWM process, without the need for iterative propagation. In particular, we show that a neural network with 3 hidden layers is able to accurately model FWM dynamics and retrieve the phase-space topology over a wide range of parameters.

We first review the theoretical description of ideal FWM dynamics. In single mode fiber, the evolution of the slowly-varying electric field envelope $\psi(z,t)$ is governed by the NLSE:

$$i \frac{\partial \psi}{\partial z} - \frac{\beta_2}{2} \frac{\partial^2 \psi}{\partial t^2} + \gamma |\psi|^2 \psi = 0, \quad (1)$$

with $z$ being the propagation distance and $t$ the time in a reference frame traveling at the group velocity. The group-velocity dispersion is $\beta_2$ and the nonlinear Kerr coefficient is $\gamma$. We consider wave mixing in the focusing regime ($\beta_2 < 0$) associated with the injection of a modulated pump wave $\psi_0$ with two sidebands at angular frequencies $\pm\omega_m$:

$$\psi(z,t) = \psi_0(z) + \psi_{-1}(z)\exp(i\,\omega_m\,t) + \psi_1(z)\exp(-i\,\omega_m\,t), \quad (2)$$

In general, the injection of such a modulated signal in an optical fiber leads to the generation of multiple additional sidebands [3, 8]. However, in the case where higher-order sidebands can be neglected as in our experiments [4], the nonlinear dynamics of the pump and two sidebands only can be described by only three coupled equations [2]:

$$\begin{cases} -i\dfrac{d\psi_0}{dz} = \gamma\left(|\psi_0|^2 + 2|\psi_{-1}|^2 + 2|\psi_1|^2\right)\psi_0 \\ \qquad\qquad + 2\gamma\,\psi_{-1}\,\psi_1\,\psi_0^* \\ -i\dfrac{d\psi_{-1}}{dz} + \dfrac{1}{2}\omega_m^2|\beta_2|\psi_{-1} = \gamma\left(|\psi_{-1}|^2 + 2|\psi_0|^2 + 2|\psi_1|^2\right)A_{-1} \\ \qquad\qquad + \gamma\,\psi_1^*\,\psi_0^2 \\ -i\dfrac{d\psi_1}{dz} + \dfrac{1}{2}\omega_m^2|\beta_2|\psi_1 = \gamma\left(|\psi_1|^2 + 2|\psi_0|^2 + 2|\psi_{-1}|^2\right)\psi_1 \\ \qquad\qquad + \gamma\,\psi_{-1}^*\,\psi_0^2 \end{cases} \quad (3)$$

This system is equivalent to a one-dimensional conservative Hamiltonian; its phase-diagram representation being preferred through the transformation of the amplitudes $\psi_k(z)$ and phases $\varphi_k(z)$ of the evolving sidebands ($k = 0, \pm1$) to canonical variables $\eta$ and $\phi$ by:

$$\begin{cases} \eta = |\psi_0|^2 / P_T \\ \phi = \varphi_1 + \varphi_{-1} - 2\varphi_0 \end{cases}, \quad (4)$$

where $P_T = |\psi_0|^2 + |\psi_{-1}|^2 + |\psi_1|^2$ is the total average power. Here, $\eta$ and $\phi$ are interpreted as the fraction of the total power in the central frequency component and the phase difference between the sidebands and the pump, respectively. The dynamics on the ($X=\eta\cos\phi$, $Y=\eta\sin\phi$) plane fully captures all the physics of this system [2, 9]. An illustration provided in Fig. 1(a) depicts phase-space portraits for various initial values of $\eta_0$ and $\phi_0$. The closed orbits obtained for $\phi_0=0$ are localized on the right-hand side of the ideal FWM separatrix orbit (dashed black line) which divides the phase space into two distinct regimes. The fiber parameters used correspond to experiment: $\beta_2 = -7.6$ ps$^2$/km and $\gamma = 1.7$ /W/km. The average power is 21.7 dBm leading to a maximal small-signal gain at the modulation frequency $\omega_m = 2\pi\,40.10^9$ rad/s.

The experimental setup we implemented is shown in Fig. 2 and is made of commercially-available telecommunications components. The key difference with respect to our setup previously described [4, 8] is that the experimentally measured values are not imprinted sequentially as new input values, and the process is not iterated. The aim here is rather to use the system to generate a data set by varying the input amplitude and phase ($\eta_{in}$, $\phi_{in}$) for various control powers $P_T$, followed by measurement of the corresponding output state ($\eta_{out}$, $\phi_{out}$), and then to train a neural network to model this input-output propagation step.

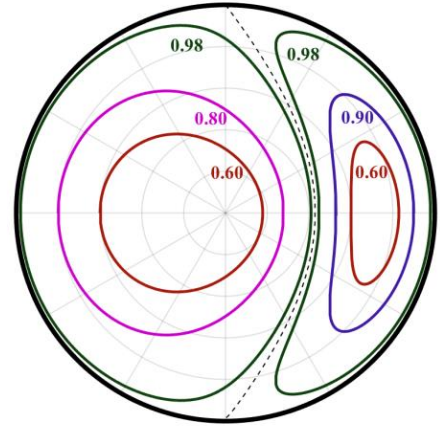

**Fig. 1.** (a) Phase space portraits for initial values of $\eta_0$ of 0.6, 0.8, 0.9 and 0.98 (red, magenta, purple and green lines, respectively). Results obtained for an input average power of 21.7 dBm are plotted for an initial phase offset $\phi_0$ of 0 or $\pi$, which appear respectively on the right and left side of the separatrix (dashed black line).

The setup is based on a continuous wave (CW) laser operating at 1550 nm, which is sinusoidally modulated using a 40-GHz phase modulator (PM) to create an equispaced frequency comb. The resulting symmetrical comb is processed using a programmable filter to tailor the three-component signal with the target $\eta_{in}$ and $\phi_{in}$, and filter out any unwanted higher order harmonics. The comb is then amplified by an erbium-doped fiber amplifier (EDFA) that delivers a constant average power $P_T$.

The nonlinear propagation takes place in a 500 m length of fiber with parameters as given above. With such a short length of fiber, detrimental effects such as Brillouin or Raman scattering are minimized, and the growth of additional sidebands remains low enough so that the framework of a degenerate four-wave interaction remains valid. The output signal is split into two channels in order to simultaneously record both spectral phase and amplitude. An optical spectrum analyzer (OSA, resolution 0.1 nm) provides directly the ratio $\eta_{out}$. The spectral phase offset $\phi_{out}$ is retrieved from the temporal delay between the central and lateral sidebands as measured by a high-speed sampling oscilloscope.

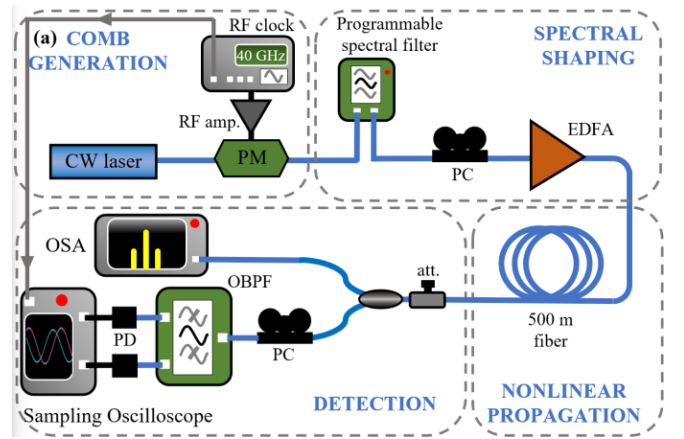

**Fig. 2.** Experimental setup. PC: polarization controller. Att: Attenuator. OBPF: optical band pass filter. PD: photodiode. RF amp: Radiofrequency amplifier. Other abbreviations are described in the text.

We recorded 6 datasets of 300 points each for power $P_T$ between 21.2 dBm and 23.7 dBm with 0.5 dBm increment. The input values ($\eta_{in}$, $\phi_{in}$) are randomly distributed in order to uniformly cover the parameter space. Figure 3 shows an example of such measurements. We then train a feedforward fully connected neural network (NN) to link the input and output states. The structure of this network is illustrated in Fig. 3(b). The input and output parameters are ($X_{in}$, $Y_{in}$, $P_T$) and ($X_{out}$, $Y_{out}$) where vectors $X$ and $Y$ contain the concatenated amplitude and phase data. The NN is made of three hidden layers involving a total of 12 neurons. The training relies on the Levenberg-Marquardt regularization back-propagation algorithm. Using 70% of our experimental data for training, the network converged in a few hundreds of epochs achieving an RMS error that is around $2.5 \times 10^{-4}$. We also benchmark our NN by comparing its predictions with the results of direct numerical simulations of Eq. (3) over a single fiber segment. In this case, an RMS deviation of $1.4 \times 10^{-2}$ is achieved for a sampling of 10 000 randomly generated inputs.

dotted lines. For each value of $\eta_0$ the NN predicted the dynamics for two values of initial phase: $\phi_0 = 0$ and $\phi_0 = \pi$, over a distance of 25 km (i.e. 50 iterations). We stress here the major difference with the sequential approach that we developed in [4] where full trajectories were sequentially recorded using iterative measurements of 50 points per trajectory, requiring 7 minutes for each trajectory. Here, once the initial dataset made up of randomly distributed points is recorded and the training stage is completed, the NN is able to quasi-instantaneously predict the evolution over the same distance without any additional experimental input. In addition, this approach results in the NN screening out naturally spurious points and thus decreases the impact of measurement noise and random drift.

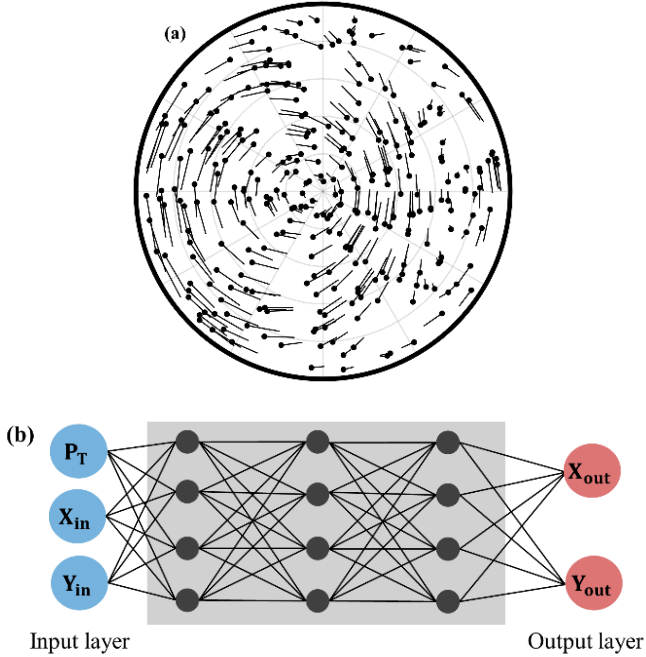

**Fig. 3. (a)** Example of one of the 6 datasets used for the training of the neural network: 300 combinations of the input parameters at a given average power (21.7 dBm). Results of each propagation are denoted by a segment with the output marked by a circle. **(b)** Neural network model used to predict the link of the output parameters with the input signal properties after a short propagation in a 500-m fiber.

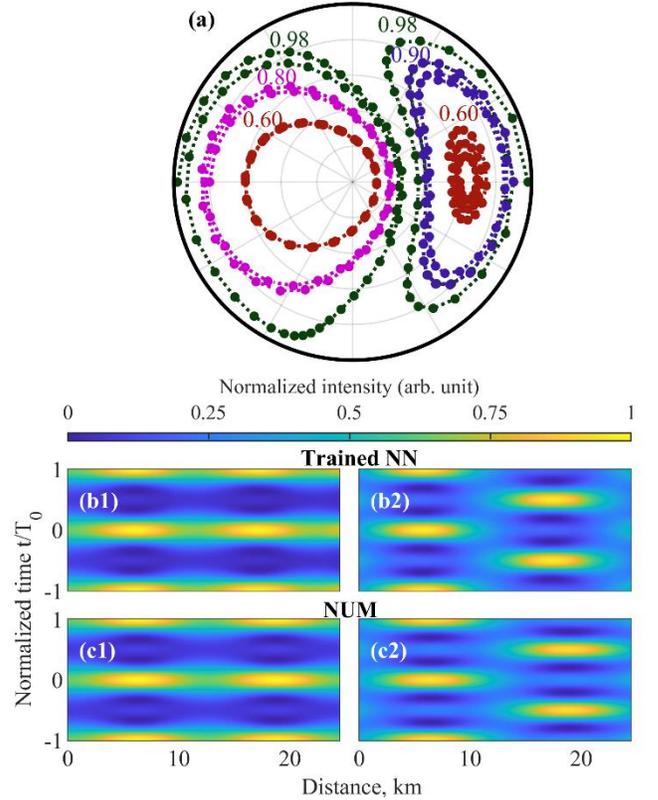

**Fig. 4. (a)** Phase space portraits for initial values of $\eta_0$ of 0.6, 0.8, 0.9 and 0.98 reconstructed by the NN for an input average power $P_0$ = 21.7 dBm (same color code as Fig. 1). Results are plotted for an initial phase offset $\phi_0$ of 0 or $\pi$. which appear respectively on the right and left sides. Predictions are made over 50 iterations. **(b-c)** Longitudinal evolution of the temporal intensity profiles predicted by the NN (panels b) or theoretically (panels c) for $\eta_0$ = 0.9, $\phi_0$ =0 and $\eta_0$ = 0.8, $\phi_0$ =$\pi$ (panels 1 and 2, respectively).

A more graphical way to confirm the accuracy of our NN is to reconstruct the complete trajectory in the phase-portrait plane. This can be easily achieved by concatenating the subsequent results: the prediction of the NN obtained after 500m propagation is used as the initial condition for the next step, therefore mimicking evolution over tens of kilometers of fibers without development of any detrimental effects. The resulting dynamics of the system at maximum gain ($P_{in}$ = 21.7 dBm) for different values of $\eta_0$ and $\phi_0$ is displayed in Fig. 4(a) with the orbits shown as circles connected by

These results yield immediate insight into the phase-space topology. The orbits obtained from the NN trained on the experimental data are in qualitative agreement with the predictions from the ideal system described in Fig. 1. The fundamental features of the ideal FWM dynamics are retrieved from these results. The trajectories are nearly closed orbits and do not intersect. We once again confirm the importance of the separatrix dividing the phase plane into two well-defined regions, with the measurements at $\eta_0$ =

0.98 providing a particularly clear indication of its location. Note that with the experimental approach of [4], due to experimental sensitivity of the measurements, propagating so close to the separatrix would often result in the latter being artificially crossed over because of measurement noise.

Significantly, with the knowledge of the spectral phase and intensity of the three interacting frequency components of the evolving field, it is straightforward to fully reconstruct the evolving intensity profiles in the temporal domain. Over a propagation distance of 25 km, Fig. 4(b) shows these results for initial values of $\eta_0$ =0.9 and 0.8 and both extremes of phase $\phi_0$ = 0 and $\phi_0$ = π. We observe typical recurrence dynamics as predicted by Eq. (3), and the case of $\phi_0$ =π also highlights a temporal half-period offset, leading, as expected by theory, to period doubling [2, 3].

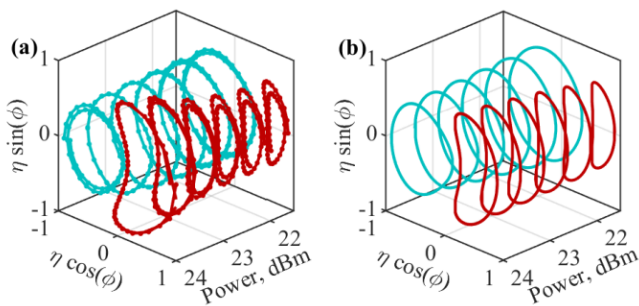

**Fig. 5.** Influence of the input average power on the phase-space portraits obtained for initial values of $\eta_0$ of 0.9 and initial phase offset of 0 or π (red and cyan color, respectively). The predictions retrieved from the neural network (panel a) are compared with the theoretical results from Eq. (3) (panel b).

As the NN is trained with data including various input average powers, it can also be used to explore unstable dynamics for other values of gain. Phase-space portraits predicted by the NN for a fixed initial value of $\eta_0$ = 0.9 and phase offset $\phi_0$ of 0 and π, and for average values between 21.5 and 24 dBm are compared with the ideal results of Eq. (3)  (Fig. 5, panels a and b, respectively). Once again, the NN-based trajectories are in good agreement with the theoretical predictions. We note how the dynamics at higher gain are associated with the change of the shape of the trajectories and the displacement of the separatrix. With increasing powers the separatrix progressively shifts: the intersection point between the separatrix and the horizontal axis $\phi$ = 0 continuously decreases. Consequently, the phase-space available for the evolution of initial conditions $\phi_0$= 0 becomes larger and larger, whereas initial conditions $\phi_0$ = π evolve in more and more restricted areas.

To conclude, we have demonstrated here that a neural network can be readily trained from experimental data to accurately model four-wave mixing dynamics in optical fiber. Based on measurements in a short fiber segment using programmable amplitude and phase, the network is shown to efficiently reconstruct the phase-space topology of a four-wave mixing over a range of parameters. In addition to providing a further example of the utility of machine learning techniques in nonlinear optics, this approach may also point to more efficient experimental implementations studying four-wave mixing over extended distances. Finally, we note that although we have focused on spectrally-symmetric initial conditions, experiments can be readily adapted to asymmetric initial sidebands with the level of asymmetry being an additional parameter of the NN [9]. Extension to other fiber-based systems involving nonlinear mixing process can also be anticipated [1].

**Funding.** OPTIMAL (ANR-20-CE30-0004); EIPHI-BFC (ANR-17-EURE-0002); I-SITE BFC (ANR-15-IDEX-0003); Région Bourgogne-Franche-Comté; Institut Universitaire de France ; CNRS (MITI interdisciplinary programs).

**Acknowledgments.** The authors also thank GDR Elios (GDR 2080).

**Disclosures.** The authors declare no conflicts of interest.

**Data availability.** The data that support the findings of this study are available from the corresponding author, CF, upon reasonable request.

**References**

1. G. P. Agrawal, *Nonlinear Fiber Optics, Fourth Edition* (Academic Press, San Francisco, CA, 2006).
2. S. Trillo and S. Wabnitz, "Dynamics of the nonlinear modulational instability in optical fibers," Opt. Lett. **16**, 986-988 (1991).
3. A. Mussot, C. Naveau, M. Conforti, A. Kudlinski, F. Copie, P. Szriftgiser, and S. Trillo, "Fibre multi-wave mixing combs reveal the broken symmetry of Fermi–Pasta–Ulam recurrence," Nat. Photon. **12**, 303-308 (2018).
4. A. Sheveleva, U. Andral, B. Kibler, P. Colman, J. M. Dudley, and C. Finot, "Idealized Four Wave Mixing Dynamics in a Nonlinear Schrödinger Equation Fibre System," Optica **9**, 656-662 (2022).
5. G. Genty, L. Salmela, J. M. Dudley, D. Brunner, A. Kokhanovskiy, S. Kobtsev, and S. K. Turitsyn, "Machine learning and applications in ultrafast photonics," Nat. Photon. **15**, 91-101 (2021).
6. S. Boscolo and C. Finot, "Artificial neural networks for nonlinear pulse shaping in optical fibers," Opt. Laser Technol. **131**, 106439 (2020).
7. L. Salmela, C. Lapre, J. M. Dudley, and G. Genty, "Machine learning analysis of rogue solitons in supercontinuum generation," Scientific Reports **10**, 9596 (2020).
8. B. Frisquet, A. Chabchoub, J. Fatome, C. Finot, B. Kibler, and G. Millot, "Two-stage linear-nonlinear shaping of an optical frequency comb as rogue nonlinear-Schrödinger-equation-solution generator," Phys. Rev. A **89**, 023821 (2014).
9. G. Cappellini and S. Trillo, "Third-order three-wave mixing in single-mode fibers: exact solutions and spatial instability effects," J. Opt. Soc. Am. B **8**, 824-838 (1991).